# ENHANCEMENT IN THE EFFECTIVENESS OF REQUIREMENT CHANGE MANAGEMENT MODEL FOR GLOBAL SOFTWARE DEVELOPMENT

Ahmed Mateen[1] and Hina Amir[2]
Department of Computer Science, University of Agriculture, Faisalabad
Corresponding Author's email: ahmedmatin@hotmail.com

**ABSTRACT**— The need for change in project requirements is necessary for every organization due to change in technology, change in government policy, and change of customer or stakeholder's requirements. Requirement Change Management (RCM) is not an easy task, especially in Global Software Development (GSD) where team members are globally distributed in different geographical location and a cultural difference is present between team members. So it becomes more difficult to manage these changes. There are a number of risks that are faced during requirement change management in global software development process. The aim of this research is to discuss these issues, tools and techniques that are being used to reduce the effectiveness of these issues in requirement change management. On the basis of these methods, propose a new model that will enhance the effectiveness of requirement change management process.

## 1. INTRODUCTION

Nowadays, in software industry, the introduction of new methods for software development is growing rapidly. Initially, the software development organizations collocated, but over time they moved into the global market [1, 2]. It provides new encouragement as global software development (GSD) is steadily gaining acceptance. Low labor costs, skilled in Personals, proximity to markets and clients, access to the latest technology by GSD software industries are some of the reasons. Similarly, geographic distance, temporal difference, likes the difference between social and cultural challenges [3] Communication, collaboration, knowledge management and knowledge sharing issues giving rise to [4]. Requirement engineering is a branch of software engineering, which is subdivided into two phases. First phase is called requirement development phase and second phase is called requirement management phase. Requirement development phase consists of activities of requirement elicitation, analysis, documentation and validation [5]. The purpose of requirement development is creating and analyzing the customer's need. Requirement management phase consists of activities of change request, change request is performed the impact analysis and on the basis of impact analysis approve or reject these changes and then implement approved changes and perform requirement traceability. It also organized a process in which requirement are arranged and saved that are required to achieve a desired goal [6].

Requirement change management is a difficult task and it became more difficult in global software development. Team members are geographically distributed in any other country of the world, where software development cost is decreased. Software companies that situated in high cost countries are trying to reduce development cost by hiring the programmer's from low cost countries from all over the world [7]. For a single organization it become difficult to provide all facilities for complex projects on single place due to lack of resources, social, commercial and economic reason and organization don't want to collaborate with other organizations that is expert in some areas of project. Therefore organizations try to seek experts in specific domain present in different geographical place. So in this way global software development becomes more important in developing software projects. Global software development process has achieved a great importance in software industry due to its several advantages that consists of decreased labor's working cost, twenty four hour development time, availability of skills persons at any time and at any place, fastest communication, increased access to customer and market and know how to new technologies. Like other technologies global software development process also has some disadvantages such as cultural and language differences, background knowledge are different, different time zones etc. These are the factors that may cause the failure of a project [8].

For decade ago, global software development's market had faced the problem of project failure. With the passage of time and with better technology, problems in global software development are minimized. However, requirement change management is a challenging task among distributed team members than collocated team members because they communicate with each other at any time and due to face to face communication. So team members are face the problem of lack of understanding of requirements and in global software development process, a number of technical and non-technical risks are involved. Technology, strategy and lack of coordination issues among systems are faced in requirement change management framework. The main causes of requirement changes are changes in customer's requirements, missing requirements, and change in technology, for functional improvement and due to change in managerial strategy [9]. Global Software Development should provide shared vocabulary to overcome language barriers [10].

In GSD requirements management, due to lack of common understanding and evolution is difficult to deal with. Therefore, knowledge of structured software development life cycle needs at any stage to deal with the changes that take place all dispersed team members should be informed by and shared. Changes often inadequately handled effect product quality and disappointing results from a technical and business team directly [11]. The requirement change management is a vital part of Software Requirements Engineering Process. It plays a vital role in the production of any software development. In fig-1 the traditional software change request process has been shown. In this process, change request should be initiated by client and after this change request manager, who is responsible to prepare a change request note. Then prioritize the requirement and determine the impact of change, next step is to resolve the conflicts among requirement. This note is forwarded to the Change Control Board (CCB), who has the authority to accept or reject the change after thorough investigation over the change impact. Then circulate this change to all concerns. This research proposes a RCM framework. The architecture of the framework suggests how it should be incorporated in Software Development Framework in order to show its effectiveness in requirement change management with respect to GSD environment.

In GSD environment developers are spread all over the world so it is a need of the current era to develop a new framework that has the ability to solve heterogeneous issues and to identify a common approach for understanding. Our objective to use requirements Model for knowledge management method is to define a mutual understanding among various stakeholders for knowledge interchange. Model describes the requirement in formal way so it also resolves the ambiguity issues of the requirements.

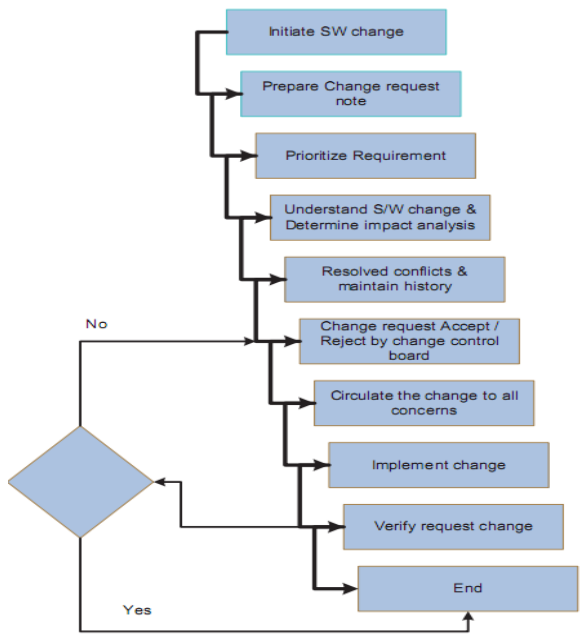

**Fig 1: Traditional Requirement Change Request Process Model [6]**

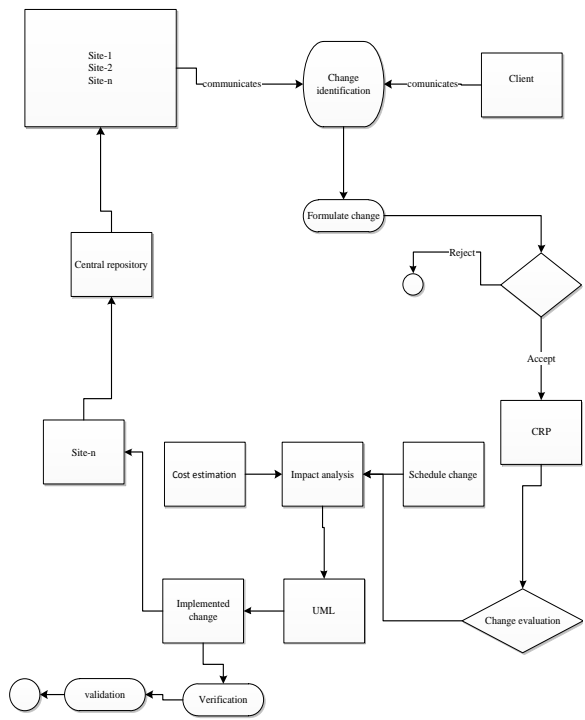

**Fig. 2: Architectural view of Proposed Requirement Change Management Model**

## 2. RELATED WORK

In the past few years different requirement change management models for global software development have been proposed. Some of these models have been studied. The brief description of these models is given below in detail such as:

In [12],Hussain and Ehsan proposed a framework that is used for managing changes in requirements for technical, industrial and business application. All changes were analyzed before being approved by the Change Control Board after analyzing the impact of change and by regulating the process for change. The complete assessment report is recorded with all changes and also documented the individual assessment report of the changes in Software Development Lifecycle process. Identified parameters are assessment report, analyses impact, control changes, record changes, proof changes by change control board, update document, change control board. In [13] authors proposed information centric and process oriented approach used for managing client requirements by using innovative framework in information management system. The ERIM framework stands for Enterprise Requirements Information Management framework and it reduces cost and time of construction project throughout project lifecycle. In [14] authors described controlled version of application and record all changes in new version of system in a database. Version control minimizes the impact of system requirement changes. Green framework based on an incremental approach and green wizard tool support reengineering and development process. This framework support reuses of software design, code. In paper [15] Chua and Vemer determined the rework effort estimation for requirement changes. The framework reduces the risks in effort estimation for rework and control changes. The framework estimated the cost of rework and analyzed the impact on requirement changes. Change control forms approved, implement change requirement, then described relationship between requirement change and rework in change management. In [16] paper Smite applied the concept of mediating partner between end user and developer. Mediating partner helps in purchasing software and support developer during development. Following problems have occurred during its working like terminology difference, lack of language skill, inconsistency in work practices, customer, employee unwillingness to collaborate and lack of version control. In paper [17] authors have been proposed a model that can be adopted by all types of software organizations. The selection of UML (Unified Modeling Language) is of great importance because proposed framework environment is dispersed. Costing is used for impact analysis and for estimating the cost respectively. Impact Analysis over change was recorded with the help Cost functions.

## 3. PROPOSED MODEL

Literature on these dynamically changing requirements can be used to minimize the ambiguity of the solution. GSD needs change in the environment, so it can be applied. Details of the proposed model are given below:

### 3.1. DESCRIPTION OF THE FRAMEWORK

We proposed a Global Software Development for "Requirement Change Management Model (RCMM)". The models suggested improving their software products to help organizations of all kinds of software that can be adopted by a framework. Evaluation of this model by feedback from expert reviews will be. RCM success of early stage in the development of the framework was to decide. To assess the quality of the induced current RCM (required Change Management) model comes from experimental studies.

Fig 2 RCM architectural framework approach shown here. The proposed model depicts the complete Requirement Change Management process in Global Software Development environment. It is a framework that could be adopted by all types of software organizations to assist them in improving their software production. The evaluation of this model will be conducted through feedback from expert reviews. The preliminary phase is the identification of change was to decide the success criteria. The motivation for estimating this criterion comes from experimental studies of existing Requirement Change Management Models. The framework is based on the quality assurance steps including Global Software Development. First of all identify the Current change request that comes from different stakeholders and users, 2) Formulate the Change, Set requirements that need to be changed, Goals and Measurements, 3) Change is accepted or rejected by the project manager. If change is rejected then process terminates. If change is accepted, then process of Change Management continues, go to next phase that is called Change Request Process. Execute Requirement Change Management Processes and Collect and Validate changed requirements 4) CRP generates change request form that is submitted to the change control board 5) Change evaluation is done by CCB board. 6) When change is evaluated then impact analysis is performed through cost estimation function, Schedule change and Unified Modeling Language is

**Table 1: Expert's Panel Opinion**

| Apparent Benefits | | Team Leader | Project Manager | Project Manager | Project Manager | Software Manager | Software Manager | Software Manager | RE Engineer | RE Engineer | QA Manager | QA Manager |
|---|---|---|---|---|---|---|---|---|---|---|---|---|
| Company without Using RCM Model | User satisfaction | P | P | O | A | A | P | A | O | N | O | A |
| | Ease of Learning | A | N | N | O | A | N | A | A | O | A | A |
| Company With Using RCM Model | User satisfaction | A | A | N | O | A | A | A | A | O | A | A |
| | Ease of Learning | A | A | O | A | N | A | O | A | A | A | A |

Agreed= A, Partially Agreed= P, Not Present= N

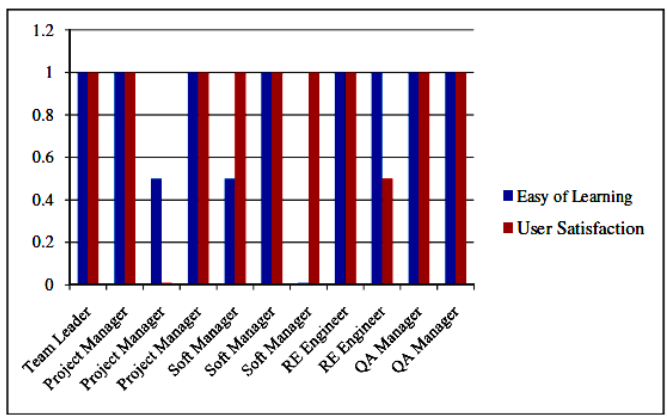

**Fig. 3: Expert Review Graph**

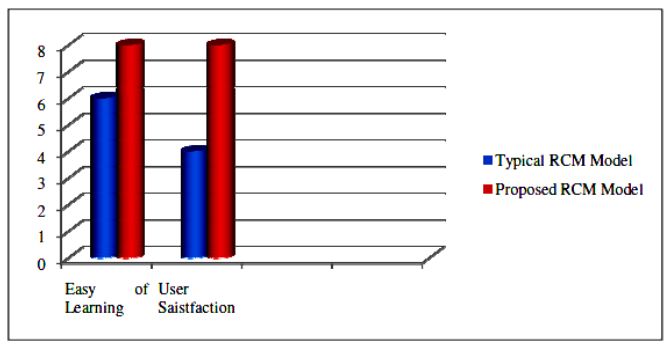

**Fig. 4: Evaluation of Two Models**

used due to Global Software Development. 7) In this step change is implemented to multiple sites and 8) verified process implementation and terminate the change. One of the key challenges is to manage teams dispersed and is trying to meet the user expectation when problems arise. Development team does not adjust the project to the client when the client is completely meaningless request. Most require the co-management model is used for software projects and accurately distributed software projects to have sufficient capacity to accommodate the needs. The excessive cost, schedule slips, and the results of most of the failed projects. Therefore, changes in the distribution of software for the management of projects to find new models are necessary. This model mainly to improve their software products for global software development has been designed. Based on facts collected from the literature and by expert's reviews, this model is being introduced.

### 4. EVALUATION
The research work will be started with a detailed literature review, Journal articles, conference proceedings, books, research reports. A questionnaire for a survey will be prepared that will consist on three basic categories technical, non-technical and attitudes (personal and behavioral) skills that are affected the requirement change management in global software development process. Scale will be used to design a questionnaire for survey. Statistical analysis will be performed on requirement change management model in global software development to check the effectiveness of proposed model [13].. GSD is following an environment that is difficult to follow a normal change management process. We apply our framework in the organization and took the expert review. Expert opinion process was used to take the opinion of RCM for Customer satisfaction and easy of learning. The expert panel that have a team leader of fifteen years of experience, three software engineers from which one has a ten year experience and two have eight year experience, three project managers that have with three, five, and seven years of experience respectively, two RE engineers having six years of experience, and two quality assurance members having five years of experience.

### 4.1. EASY OF LEARNING
According to the expert's opinion this model is easy to understand.

### 4.2. USER'S SATISFACTION
Our model would be helpful for the software industry in general in expert's opinion. Experts that have used other models noted that our framework was giving better outcomes. But according to some experts, our model is need to evaluate by more case studies for better results in GSD to implement large scale software industries. The results of our evaluation is the opinion of various experts as they agreed, partially agreed and did not agree, which is shown in Table 2. Expert opinion of a graph is also shown in Fig 3. Our model and the performance difference between the RCM processes are shown in Fig. 4, and it is clear that the proposed model is perfect for organizations.

Based on the initial evaluation, the experts understand the needs of the management model are changing and the change management requirements may well think that. However, other companies to meet their requirements for change management framework may need to use.

### 5. CONCLUSIONS
The purpose of this study is to support and formalize the success in the global software development environment that will require changes in management measurement framework. We use knowledge for knowledge management Framework for the assessment of the anticipated expert judgment is used to review a set of research questions and research reports are prepared to deal with. The main advantage of this model is that requirement changes can occur at any phase of system development. In GSD due to communication issues RCM is difficult to manage. The proposed framework help to improve the understanding of roles, work of art and behavior involved in the GSD particularly from the perception of change management system.